\documentstyle{memsait}

\newcommand{\kap}[1]{Sect.\,\ref{#1}}   
\newcommand{\abb}[1]{Fig.\,\ref{#1}}

\newcommand{\spr}{\mbox{$s$-process}}  
\newcommand{\etal}{et al.\ }
\newcommand{\msun}{\, {\rm M}_\odot}
\newcommand{\nat}[2]{#1 \cdot 10^{#2}}  
\newcommand{\hevi}{^{4}\textrm{He}}
\newcommand{\czw}{^{12}\textrm{C}}
\newcommand{\nvi}{^{14}\textrm{N}}
\newcommand{\nfu}{^{15}\textrm{N}}
\newcommand{\ndr}{^{13}\textrm{N}}
           
\newcommand{\oac}{^{18}\mathrm{O}}           
\newcommand{\cdr}{^{13}\textrm{C}}
\newcommand{\fne}{^{19}\textrm{F}}
\newcommand{\cvi}{^{14}\mathrm{C}}         
\newcommand{\fese}{^{56}\mathrm{Fe}}         
\input epsf.sty
\begin{opening}
\title{FORMATION OF THE NEUTRON DONOR $^{\mathsf{13}}$C 
IN AGB STARS BY OVERSHOOT AND ROTATION} 
\author{FALK HERWIG$^1$ AND NORBERT LANGER$^2$}
\institute{$^1$Universit\"at Potsdam, Potsdam, Germany (fherwig@astro.physik.uni-potsdam.de)\\
$^2$Universitet Utrecht, Utrecht, The Netherlands (N.Langer@astro.uu.nl)}
\date{} 
\end{opening}

\begin{document}

\oddpagefooter{}{}{} 
\evenpagefooter{}{}{} 
\ 
\bigskip

\begin{abstract}
Observations of heavy elements in Red Giant stars 
clearly show that low-mass AGB stars can provide a nucleosynthesis site
of the \spr. Stellar evolution models produced over the last years indicate 
that radiative burning of $\cdr$ between succeeding 
thermal pulses in low-mass AGB star models 
may indeed provide the neutrons for the \spr. 
However, although it seems clear that some mixing
between the proton-rich envelope and the carbon-rich core
may lead to the production of $\cdr$, the physical mechanism
responsible for such mixing is not yet unambiguously identified.

We present stellar model calculations which include mixing due to 
\emph{overshoot} and \emph{rotation}. Overshoot, with a
time-dependent and exponentially decaying efficiency, leads to a
partial mixture of protons and $\czw$ during the third dredge-up, when
the envelope convection zone reaches deep into the core. According to
the depth-dependent ratio of protons and $\czw$, a small $\cdr$ pocket
forms underneath a $\nvi$-rich layer. Overshoot 
does not allow for any mixing after the envelope convection zone
retreats at the end of the third dredge-up after each pulse. 

Rotation introduces mixing driven by
large angular velocity gradients which form at the envelope-core
interface in AGB stars, in particular after a thermal pulse. This
leads to partial mixing after a pulse, 
as in the case of overshoot. However, both mechanisms differ 
during the interpulse phase. Rotation continues to mix the
region of the $\cdr$-pocket with a diffusion coefficient of
$log D \sim 2 \dots 3 \mathrm{cm^2 s^{-1}}$. This does not only spread
the $\cdr$-pocket, but also the more massive $\nvi$-rich layer, and
finally leads to mixture of both layers. 
By the time when the temperature there has risen to 
about $9\,10^7\,$K and neutron production sets in, 
the $\nvi$ abundance exceeds
the $\cdr$ abundance by a factor of $5 \dots 10$. We analyze
the role of $\nvi$ as a neutron poison by considering
the recycling of neutrons via
$\nvi(\mathrm{n},\mathrm{p})\cvi$ 
and $\czw(p,\gamma)\ndr (\beta^+)\cdr$ qualitatively. 
We find that as long as $X(\nvi) \ll
X(\czw)$, the \spr\ will still be possible to occur under radiative conditions.
\end{abstract}

\section{Introduction}
Observationally, AGB stars show a clear signature of ongoing
\spr-nucleosynthesis. Among the evidence is the  detection of
the radioactively unstable element Tc (Merill 1952).  
Its longest living isotope, $^{99}$Tc, has a half-life of
$2.1\cdot10^5\mathrm{yr}$ which clearly exceeds the life time of a thermally
pulsing AGB star. Thus, any detected Tc must have been produced \emph{in
  situ}. More evidence comes, e.g., from the  correlation of 
heavy element abundances (like the index [hs/ls]) with metallicity (Busso
\etal 1999) or with the fluorine abundance (Jorissen \etal 1992). Another
source of information are pre-solar dust grains from meteorites. Most
of the silicon carbide and corundum grains formed in the winds of red
giants and AGB stars (Hoppe \& Zinner, 2000). The grains allow a
determination of the thermodynamic conditions of the $s$-process via
precise measurements of the branching ratios of, e.g., krypton, which
supports AGB stars as \spr-site.

The current scenario of the \spr\ in AGB stars is closely connected to
the alternating mixing and nuclear burning events which characterize
the thermal pulse cycle. For any \spr\ in AGB stars, the third
dredge-up is a necessary ingredient. It leads to a contact layer where
the H-rich envelope and the carbon-rich intershell region can exchange
some material and form a partially mixed zone. Gallino \etal\,(1998)
have assumed that a proton-enriched layer forms at the bottom of the
convective envelope at the end of the third dredge-up phase. 
In the further course
of the TP cycle, this carbon-rich layer allows the formation of
excess $\cdr$. Due to the assumed low proton abundance,
 no $\nvi$ was produced and therefore the role of this isotope as a
 neutron poison could
 be neglected. 
This so-called $\cdr$-pocket is the indispensable prerequisite 
for the radiative
neutron-capture nucleosynthesis during the interpulse phase in
AGB stars (Straniero \etal 1995, Gallino \etal 1998). 
This paper presents two physical candidate processes for
the formation of such a $\cdr$-pocket  at the envelope-core
interface of low-mass AGB stars: rotation and overshoot. 

In \kap{sec:ov} we review the properties of the overshoot mechanism as
originally proposed for AGB stars by Herwig \etal (1997). Following 
\kap{sec:rot} gives updated account of AGB models with rotation. A
comparison and discussion is presented in \kap{sec:con}

\section{Overshoot models}
\label{sec:ov}
\begin{figure}
\epsfysize=8cm 
\hspace{0.8cm}\epsfbox{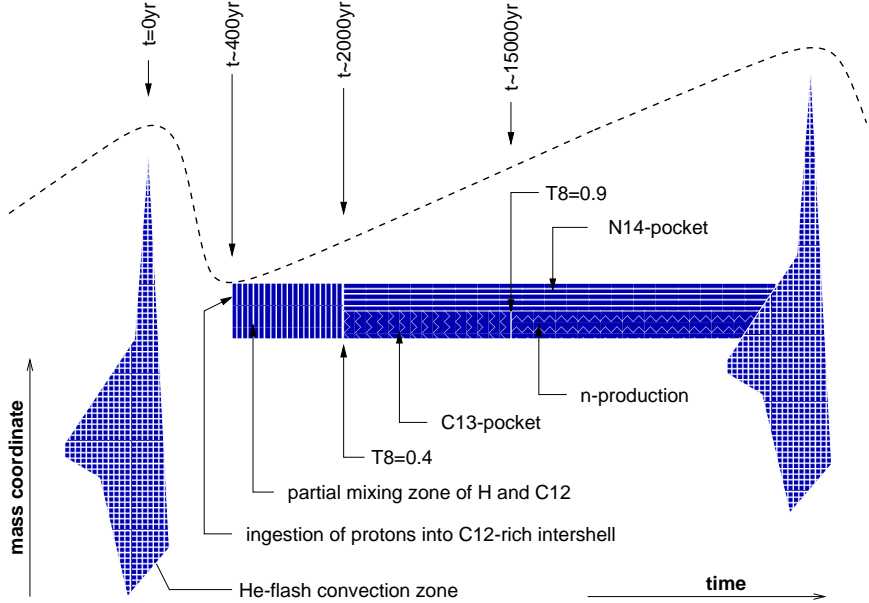} 
\caption[h]{\label{fig:schematic}
Schematic representation of the radiative $\spr$ during the interpulse
phase of TP-AGB stars with a $\cdr$ pocket due to overshoot
(instantaneous ingestion of protons into the $\czw$-rich intershell). 
The two small-checked zones represent the convective He-flash region
while the other shaded regions denote the partial mixing zone during
different stages: coexistence of protons and $\czw$ (vertically
hatched region), $\nvi$-pocket (horizontally hatched), $\cdr$-pocket
(vertical zigzag  hatch), burning of $\cdr$, n-production (horizontal
zigzag  hatch). Note, that in the overshoot picture no mixing of the
n-production layer is allowed after the retreat of the envelope
convection zone.} 
\end{figure}
The relevance of overshoot in AGB stars for the neutron source problem
has been studied by Herwig \etal (1997, 1999). 
We assumed that the
depth dependence of the mixing due to convective 
overshoot can be described by an exponentially decaying efficiency. 
This behavior is qualitatively suggested by many
hydrodynamical simulations of convection under various conditions.
Freytag \etal (1996) found that the particle spread due to
successive convective downdrafts and plumes which reach beyond the convective
boundary can be described by a diffusion-like process. Moreover, 
the velocity field in the stable region declines exponentially.
Accordingly the material mixing in the overshoot region can be  
approximated by a diffusion coefficient which decreases exponentially
from a MLT value inside the convective boundary. Herwig (2000) presents
the evolution of the abundance profiles of the core-envelope interface
during an interpulse 
period and shows snapshots (Fig.\,4) after the ingestion of protons into
the intershell layer, after the initial proton capture phase where the
$\cdr$-pocket is formed and after consumption of $\cdr$ due to
$\alpha$-capture. The situation is sketched in \abb{fig:schematic}. The
main difference to the previously and in detail studied profile by
Gallino \etal (1998) is the continuous variation of both the proton- and the
$\czw$-abundance and thus the p/$\czw$ - ratio. This ratio determines
the $\cdr$/$\nvi$ - ratio after consumption of hydrogen in the
respective layer (Fig.\, A.1 in Herwig 1998). In the overshoot case we
found that the partial mixing zone consists of two adjacent regions
after the end of proton captures. Further outward a region with $\nvi$
as the most abundant element forms. All $\czw$ is transformed into
$\nvi$ which reaches a mass fraction of $\sim 40\%$ in models where
overshoot has been applied to all convective boundaries. Below this
zone the actual $\cdr$-pocket forms where the proton abundance is too
low to continue p-captures after the production of $\cdr$. Here,
$\nvi$ is only of very small abundance. Preliminary \spr\ studies of
this situation have shown that qualitatively the reproduction of a
solar-like distribution of the main $\spr$-component is
possible (Gallino \& Herwig, 1998). 
However, it should be emphasized that the efficiency of
overshoot, hence the velocity scale height is a free
parameter. Currently we can not constrain the velocity scale height at
the bottom of the convective envelope of AGB stars during the
dredge-up. For most of the evolutionary phases we used a value of
$f=0.016$ ($H_{\mathrm{v}}=f\cdot H_{\mathrm{p}}$, $H_{\mathrm{p}}$
pressure scale height) which reproduces the
width of the main-sequence of intermediate mass stars. However, this
value produces only a pocket 
of integrated $\cdr$ of $3\cdot 10^{-7}\msun$. This small amount of
$\cdr$ leads to fairly small enhancement factors. With a larger
overshoot efficiency ($f=0.128$) we found enhancement factors of $\sim
100$ for the main component.

Another interesting property of the \spr\ in low-mass AGB stars is now
emerging from high precision measurements of pre-solar grains (A.\
Davis, this volume). Apparently these new observational data require
for a model sequence of given initial mass and metalicity a spread in
neutron exposures in the range up to a factor of up to 10. Such a
spread can not currently be reproduced by the overshoot scenario
(however, see \mbox{Lugaro \& Herwig, in prep.}). 
Although the absolute amount
of $\cdr$ is likely to be dependent on mass and metallicity for a given
velocity scale height of the overshoot, we can not assume that stars
even of different mass and metallicities have so greatly different
overshoot efficiencies at this convective boundary as to account for
the spread required by the pre-solar dust grains. It does seem
possible that the overshoot decay has a different rate at different
convective boundaries and during different evolutionary
phases. Freytag \etal (1996) found $f = 0.25 \dots 1$ for the shallow
surface 
convection zones of A-stars and white dwarfs. Such a large value is
certainly not appropriate in the adiabatic core regions and can in fact
be excluded for the main-sequence convective core of intermediate mass
stars. However, within the physical picture of overshoot the
efficiency of this mechanism should be of the same order within a
small range in similar evolutionary conditions.

\section{Rotation models}
\label{sec:rot}

Rotationally induced mixing is another  physical process which might
be responsible for the formation of the neutron donor $\cdr$ at the
envelope-core interface of low-mass AGB stars. In particular, it may
introduce a statistical signature on the \spr\ of otherwise identical
stars due to a spread of initial rotational velocities. Here we present
more detailed properties of the AGB models by Langer \etal (1999) with
respect to the formation and destruction of $\cdr$.

\subsection{Input physics and stellar models}
\label{sec:inputphysics}

The input physics is the same as in Heger \etal (2000).
The 1D-hydrodynamic stellar evolution code  contains angular momentum
as an additional state variable, and the  effect of centrifugal force
on stellar structure. Rotationally induced transport of angular
momentum and chemical species are taken into account according to
Eddington-Sweet circulations, the Solberg-Hoiland and the
Goldreich-Schubert-Fricke instability, as well as   dynamical and
secular shear instabilities. The $\mu$-gradient acts as barrier for
rotationally induced mixing ($f_\mu = 0.05$) and convection
(Ledoux-criterion, semi-convection). Nucleosynthesis is treated by a
network 
with pp-chain, CNO-, NeNa- and MgAl-cycles, He-burning including
$(\alpha,n)$ reactions and the inclusion of $(n,\gamma)$  reactions.
Here we also discuss the additional effect of the
$\nvi(\mathrm{n},\mathrm{p})\cvi$ reaction.

The stellar model sequence had an initial mass of $3\msun$ and initial
equatorial rotation velocity of $250 \mathrm{km/s}$ and was evolved to
the AGB with mass loss.
Contrary to the models with overshoot dredge-up sets
in fairly late and remains week. The TP cycle described below is the 
25$^{\mathrm{th}}$ (M$_{\mathrm{c}}=0.746\msun$) and the dredge-up just reaches 
into the carbon-rich intershell. Sufficient carbon star formation can
not be expected from these models.

\subsection{Rotationally induced mixing, formation and evolution of
the  $\cdr$-pocket}
\label{sec:rotmix}
\begin{figure}
\epsfysize=6cm 
\hspace{1.8cm}\epsfbox{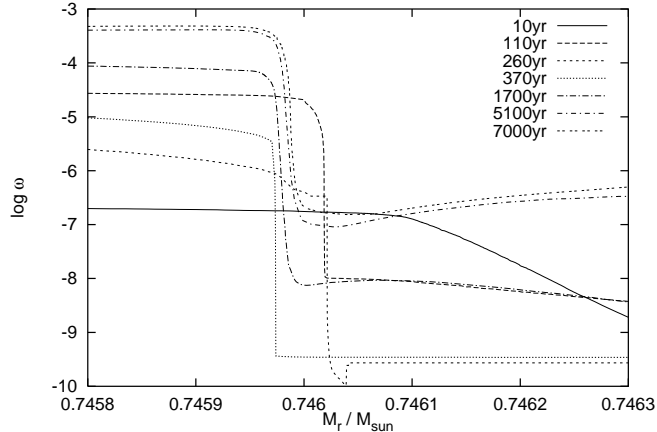} 
\caption[h]{Evolution of angular velocity during the
  25$^{\mathrm{th}}$ thermal pulse and
  the following first third of the interpulse phase in the stellar
  interface region of envelope and core (cf.\ Fig.\,3 in \mbox{Langer \etal
  1999}). The left part of the diagram below the mass coordinate $\sim
  0.746\msun$ corresponds to the top of the intershell region, the right
  part shows the bottom of the envelope. The time $10 \mathrm{yr}$
  corresponds to the onset of the He-flash just before the luminosity
  peak is reached. \label{fig:angular-vel}}
\end{figure}
In a very generalized sense rotationally induced mixing occurs where
large angular velocity gradients develop. In addition, convective
transport of angular momentum enforces close-to-ridged
rotation. Conservation of angular momentum leads to a decrease
(increase) of angular velocity in expanding (contracting)
layers. These mechanisms can be seen at work during a thermal pulse
and the subsequent interpulse phase in \abb{fig:angular-vel}. At the
onset of the He-flash ($10\mathrm{yr}$) the pulse-driven convection
zone is developing but has not yet reached the interface region  shown
in \abb{fig:angular-vel}. Accordingly the angular velocity profile
runs still smoothly in the interface region. At ($110\mathrm{yr}$) -
shortly after the maximum extension of the He-flash convection zone
into the interface layer - convection has lifted the angular velocity
below the interface and left behind a prominent step-like edge. At the
same time the descending envelope convection bottom decreases the
angular velocity above the interface which leads to an even larger
angular velocity gradient. About $150\mathrm{yr}$ later the angular
velocity in the upper core region decreases due to the expansion of
that layer following the energy ingestion further below during the
thermo-nuclear runaway of the He-shell. However, during this phase the
maximum angular velocity gradient is reached and the velocity
difference in the top layer of the core and the bottom layer of the
envelope amounts to about five orders of magnitude. 

During the further interpulse evolution the angular velocity generally
increases because contraction resumes in between the thermal
pulses. The angular velocity step, however, decreases only slowly. At 
$7000\mathrm{yr}$ -- half way from $\cdr$-formation to -destruction --
the angular velocity step at the interface still comprises almost four
orders of magnitude. Thus, rotationally induced mixing at the
interface of core and envelope after a thermal pulse and during the
interpulse is \emph{ongoing}. This is an important difference to the
overshoot picture where no mixing occurs during the interpulse phase
in the interface layer and the interpulse nucleosynthesis is \emph{static}.
 
\begin{figure}
\epsfysize=7cm 
\hspace{.5cm}\epsfbox{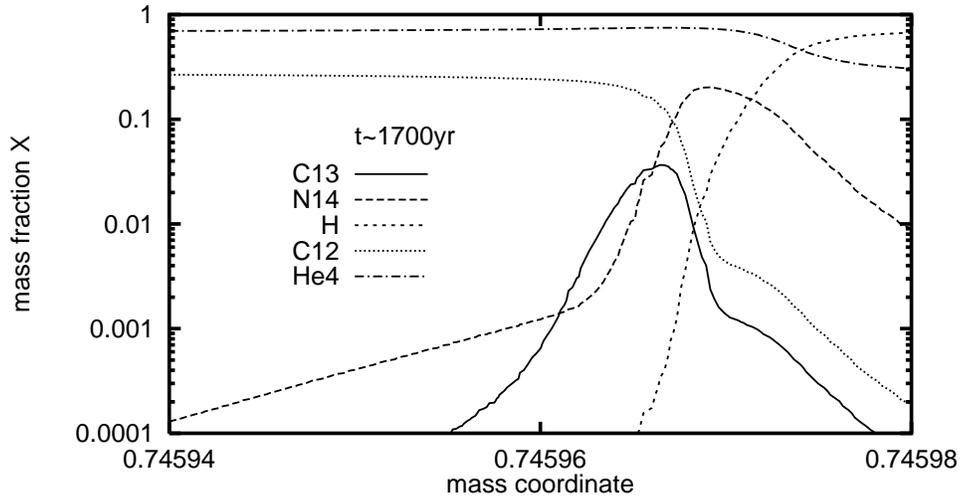} 
\caption[h]{\label{fig:formation} Abundance profiles when the $\cdr$ pocket has formed at the 
interface of core and envelope from the partial mixing of protons and 
$\czw$ due to rotationally induced mixing.}
\end{figure}
Although the physical principles of mixing in the overshoot and the
rotational case are very different the abundance profiles at the time
after the initial proton-capture phase in the partial mixing zone are
quite similar. The case where partial mixing is caused by rotation
is displayed in \abb{fig:formation}. Clearly visible are two neighboring
zones where $\cdr$ and $\nvi$ respectively are dominating. Another
similarity to the overshoot case is the negative curvature of the
proton profile in the overshoot region just before the onset of proton
captures (a few hundred years before the situation shown in
\abb{fig:formation}). A difference to the overshoot case is the larger
extent of the partial mixing zone and the tail of $\nvi$ which reaches
deep into the intershell even below the $\cdr$-pocket itself. This
$\nvi$ tail is due to the fact that the dredge-up is fairly weak and
(by coincidence?) just reaches the top of the intershell region which
has been homogenized by the previous thermal pulse. A pronounced
region of rotationally induced mixing which forms shadow-like in the
upper part of the receding He-flash convection zone (cf.\ Fig.\,3 in
\mbox{Langer \etal 1999}) causes exchange of material with the
$\nvi$-rich He-buffer zone, left over by H-shell burning during the
previous intershell phase.

\begin{figure}
\epsfysize=7cm 
\hspace{1.4cm}\epsfbox{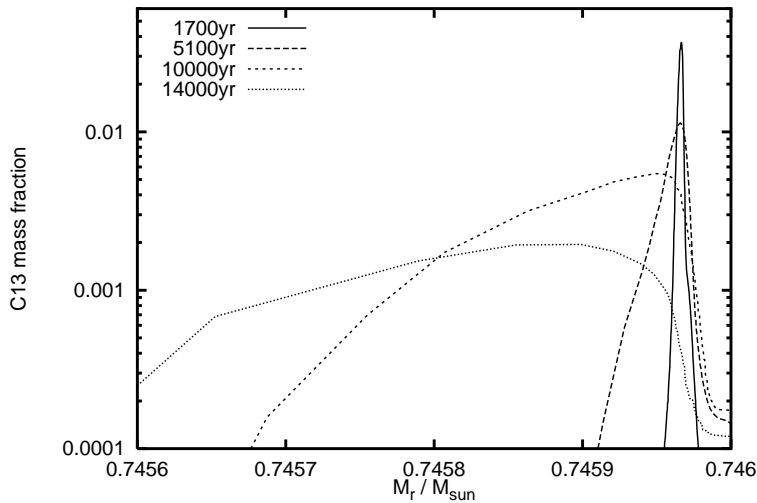} 
\caption[h]{\label{fig:C13-select} 
Evolution of the $\cdr$-pocket after its formation during the first
part of the interpulse period. The change of the profile from the initial 
formation at $\sim 1700 \mathrm{yr}$ until the onset of $\alpha$-captures of 
$\cdr$ at about $10000 \mathrm{yr}$ is due to rotationally induced mixing. 
Thereafter, 
both mixing as well as nuclear burning (neutron production) are responsible
for the modification of the abundance profile. 
}
\end{figure}
During the following phase the $\cdr$ pocket is subject to continuous
modification and the evolution is shown in
\abb{fig:C13-select}. It takes about $\sim 8000\mathrm{yr}$ from the
end of $\cdr$-formation (at $\sim 1700 \mathrm{yr}$) to the onset of
neutron production ($\sim 10000 \mathrm{yr}$)\footnote{All 
  timescales given are strongly dependent on core mass, and
  e.g.\ considerably longer for lower core masses.}. The abundance
profile change during this phase shows a broadening and lowering of
the peak $\cdr$-abundance and is due only to rotationally induced
mixing. Even after the $\alpha$-capture reactions on $\cdr$ have
started, mixing remains an important contribution to the abundance
profile modification (\abb{fig:C13-select}, $t=14000\mathrm{yr}$).

\subsection{Recycling of neutrons}
\label{sec:recycl}
\begin{figure}
\epsfysize=7cm 
\hspace{1.5cm}\epsfbox{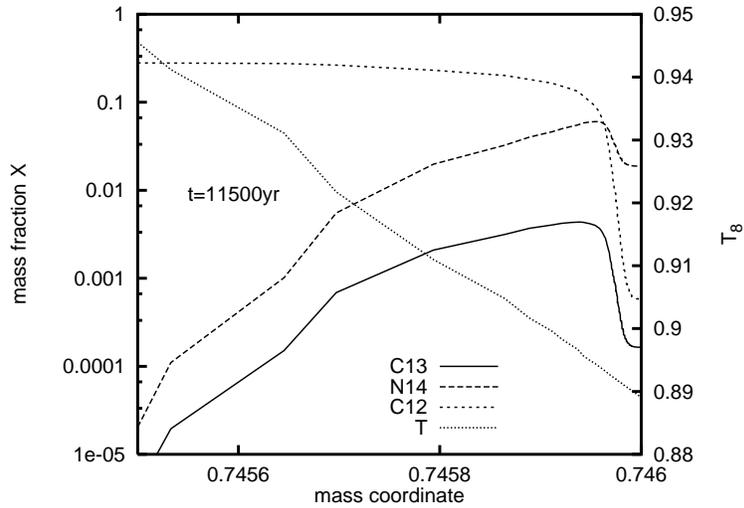} 
\caption[h]{\label{fig:N14cover} At $t=11500\mathrm{yr}$ the neutron production
is close to the maximum. Due to continuous rotationally induced mixing the 
large $\nvi$-pocket (\abb{fig:formation}) has intermingled with the
smaller $\cdr$-pocket. Due to this continuous mixing $\nvi$ is more
abundant than $\cdr$ in the region of the pocket. Note, that for the
profiles shown here, the $(\mathrm{n},\mathrm{p})$ of $\nvi$ has been
neglected. The influence of the neutron  recycling initiated by this
reaction is described in the text.
}
\end{figure}
The broadening of the $\cdr$-profile and the lowering of the peak
$\cdr$-abundance (which reaches $10\%$ in the overshoot case if the
extra mixing is applied to all convective boundaries, Herwig 2000) is a feature
which at first glance is advantageous for the \spr. The neutron
density is reduced and the exposure of neutrons is distributed over a
larger variety of seed material. However, the role of $\nvi$ is of
great importance if rotationally induced mixing is efficient. Not only
the $\cdr$-pocket is broadening as shown in \abb{fig:C13-select} but the
prominent $\nvi$-pocket shown in \abb{fig:formation} shares this
fate. The situation at about the peak neutron production is shown in
\abb{fig:N14cover}. The $\cdr$-pocket is entirely wrapped into the
broad $\nvi$-pocket. This $\nvi$ causes two problems for an efficient
\spr. The neutron-density is reduced because
$\nvi(\mathrm{n},\mathrm{p})\cvi$ is 
efficiently \emph{competing} with the n-capture of $\fese$ and
other heavy elements. Moreover, the  released protons from the
$(\mathrm{n},\mathrm{p})$ 
reaction \emph{destroy}  $\cdr$ and thereby  reduce the n-source.
$\nvi$ adds another efficient n-sink, reduces the available amount of
$\cdr$
and thereby weakens the n-source.

\begin{figure}
\epsfxsize=8cm 
\hspace{2.4cm}\epsfbox{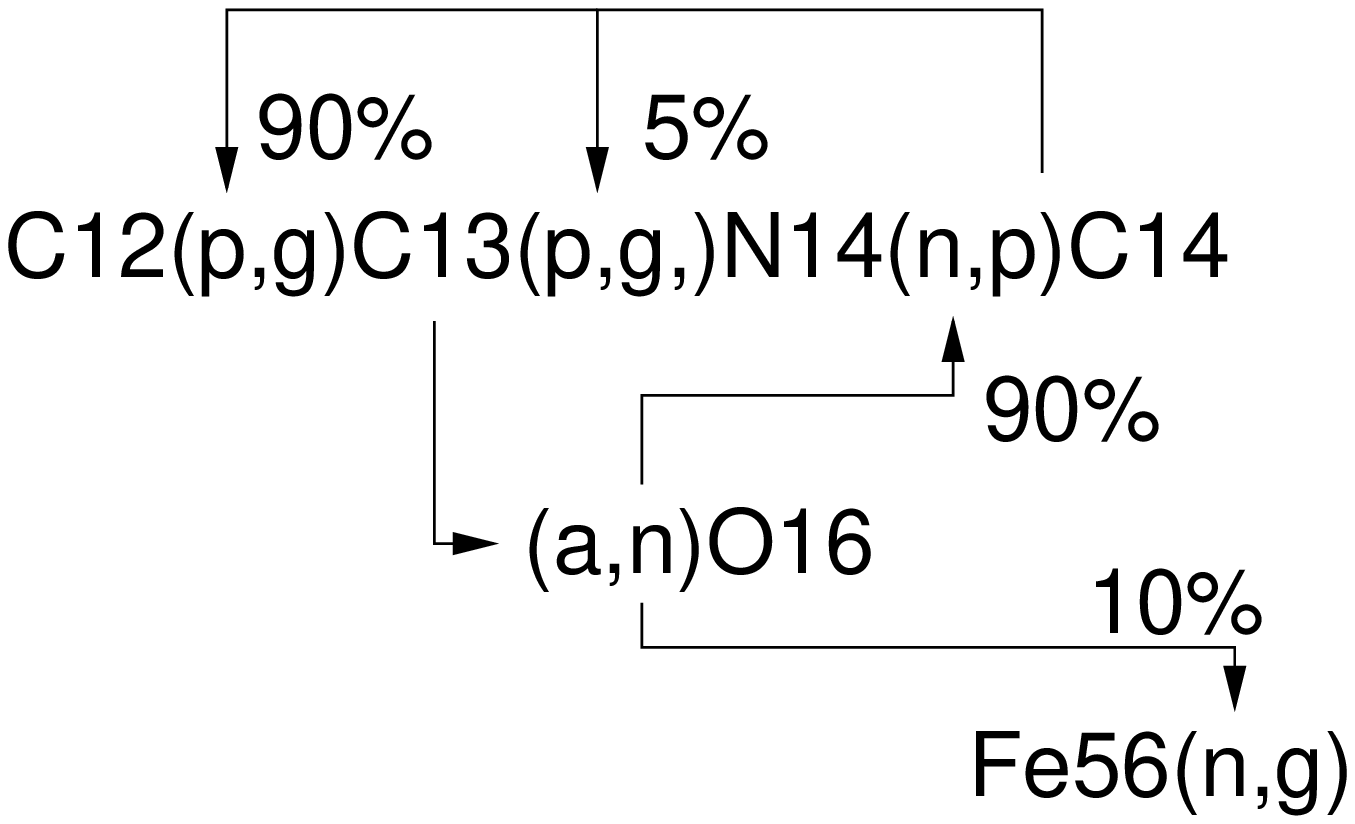} 
\caption[h]{\label{fig:recycle} Schematic representation of the main
  reactions involved in 
  the recycling of neutrons. The
  efficient $(\mathrm{n},\mathrm{p})$ reaction of $\nvi$ acts as the
  main competitor for n-captures of the Fe-seeds. If the $\czw$
  abundance is large compared to the $\nvi$ abundance more neutrons
  will escape the recycling via Fe-capture than protons via capture by
  $\cdr$. In that case the \spr\ is possible. The percentages indicate
  the branching ratios for the particular situation discussed in the text.}
\end{figure}
It is beyond the purpose of this paper to present a full nuclear
network analysis of the neutron-recycling initiated by the efficient
$(\mathrm{n},\mathrm{p})$ reaction of $\nvi$. However, some
qualitative considerations show that the \spr\ is not necessarily
prohibited by rotationally induced mixing. 
Typical conditions during n-production in the interpulse phase are given
by T$_8=0.9$, $\rho = 3700 \mathrm{g/cm^3}$, $X(\cdr)=\nat{3}{-3}$, 
$X(\nvi)=\nat{2}{-2}$, $X(\czw)=\nat{2}{-1}$, $X(\hevi)=0.7$ and 
$X(\fese)=\nat{2}{-3}$ (mass fractions). A rough estimate of the
effect of $\nvi$ as a 
neutron sink competing with iron is given by
\begin{displaymath}
  \frac{<\sigma v>_{\fese,n} Y_{\fese}}
       {<\sigma v>_{\nvi,n} Y_{\nvi}}
  \simeq 0.1
\end{displaymath}
which ignores other neutron poisons as well as other heavy elements,
in particular iron peak elements. Thus, under the given conditions of
a partial mixing zone due to rotationally induced mixing, about 9
times more neutrons are captured by $\nvi$ than compared to the
number captured by $\fese$. The effect of $\nvi$ as a source of
protons that destroy the neutron source $\cdr$ can be estimated by the
expression
\begin{displaymath}
  \frac{<\sigma v>_{\cdr,p} Y_{\cdr}}
       {<\sigma v>_{\czw,p} Y_{\czw}}
  \simeq 0.05
\end{displaymath}
where small leakages of protons  to $\oac(p,\alpha)\nfu$ ($\sim 1\%$) and 
$\nfu(p,\alpha)\czw$ ($\sim 2\%$) are neglected.

\begin{figure}
\hspace{2.4cm}\epsfbox{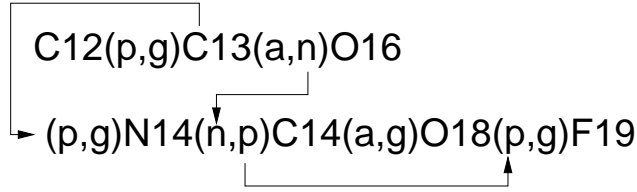} 
\caption[h]{\label{fig:f19} 
Possible radiative production of $\fne$ during the interpulse in the
partial mixing zone of a rotating AGB star. 
}
\end{figure}
The main reactions of the neutron recycling is shown in
\abb{fig:recycle}. The percentages given in the graph represent the
ratios derived above. Thus, they are specific to the case discussed
here. The chances for the iron seeds to get hold on neutrons are better
the higher  the $\czw$ abundance and the lower the $\nvi$
abundance. The larger the ratio of neutrons captured by $\fese$ to
protons captured by $\cdr$ the higher is the \spr\ efficiency because
the relative loss of protons by $\cdr(p,\gamma)$ smaller than
fraction of neutrons captured by heavy elements.
Recycling occurs on timescale of $\alpha$-capture of $\cdr$ ($\sim
13000\mathrm{yr}$ at beginning of n-production and decreasing to $\sim
1000\mathrm{yr}$ towards end of interpulse phase) and it thus leads to a
certain delayed neutron exposure. $\nvi$ as a n-poison is less
effective compared to the situation in the $\nvi$ pocket of an overshoot
partial mixing zone because the $\czw$ abundance is larger and the
$\nvi$ abundance smaller. 
Finally, the  $\nvi(n,p)$ reaction releases protons at high
temperatures and thus AGB models with rotation may be able to produce at
the same time heavy elements and $\fne$ as shown in \abb{fig:f19}.

\section{Conclusions}
\label{sec:con}

In this paper we have discussed in some detail the evolution 
of the abundance of
the neutron donor $\cdr$ according to stellar models including overshoot
and rotation. Although both processes give promising results,
neither of the two classes of models was found to be
free of problems and unsolved aspects. 

Rotation can produce a $\cdr$-pocket, which, however, is mixed with
a $\nvi$-rich layer during the interpulse phase. This makes
the consideration of $\nvi$ as neutron poison important, although,
as shown above, a larger $\nvi$ than $\cdr$ abundance 
does not automatically imply that the \spr\ is choked off.
In any case, the models with diffusive overshoot are free of this
complication as no mixing occurs in the $\cdr$-pocket in between
two thermal pulses. 
This situation has also the advantage that it 
can be accurately modelled by a series of one-zone nuclear
network calculations for the conditions in various layers of the
interface region of envelope and core. 
In order to
model the $s$-process in rotating AGB stars, 
nucleosynthesis calculations in many layers covering the interface
region, must be alternated with time-dependent mixing according to the
rotationally induced mixing efficiency. 

The most important next step will be
to quantify the neutron recycling during the interpulse phase in
rotating AGB stars.
With respect to overshoot, the main problem remains the unknown
efficiency of this process. 
Current models indicate that the \spr\ efficiency 
reaches acceptable levels within the lifetime of an AGB star only if an
overshoot efficiency above that obtained fitting the
width of the main sequence band
($f=0.016$; cf. Herwig \etal 1997, Herwig 1998)
is chosen during the dredge-up phase at
the bottom of the envelope convection. 

At present, we can only conclude that both, rotation and overshoot,
have the potential to play an important role in the \spr\ nucleosynthesis
in AGB stars, but we are  unable to quantify the relative importance of
these mechanisms for the production of the neutron donor $\cdr$ yet. 
However, in view of the results presented above, it appears not
unlikely that both processes act in fact simultaneously. A study
of stellar model sequences including 
the combined consideration of these two 
processes will be done in a future investigation.
 
\acknowledgements This work has been supported by the \emph{Deut\-sche
  For\-schungs\-ge\-mein\-schaft, DFG\/} (La\,587/16). We would like
to thank Roberto Gallino and Maria Lugaro for important and helpful
discussions.


\end{document}